\documentclass{osa-article}
\usepackage[percent]{overpic}
\journal{osajournal}

\articletype{Research Article}
\begin{document}

% \title{Dressed Four wave mixing in hot vapours described by microscopic model}

\title{Spectral control of quantum correlations in Four wave mixing using %trough
dressing fields}

\author{A. Monta\~na Guerrero\authormark{1}, R.L.Rincon Celis,\authormark{1},  M. Martinelli and H. M. Florez\authormark{1,*}}

\address{\authormark{1}Instituto de F\'{\i}sica, Universidade de S\~ao Paulo, 05315-970 S\~ao Paulo, SP-Brazil
% \authormark{2}Publications Department, The Optical Society (OSA), 2010 Massachusetts Avenue NW, Washington, DC 20036, USA\\
% \authormark{3}Currently with the Department of Electronic Journals, The Optical Society (OSA), 2010 Massachusetts Avenue NW, Washington, DC 20036, USA}
}
\email{\authormark{*}hans@if.usp.br} %% email address is required



\begin{abstract}
We present a % the first 
microscopic description of dressed four wave mixing (4WM) demonstrating spectral control of quantum correlations. Starting from a  
double $\Lambda$ model for a single pump 4WM, we include a dressing field coupling the excited level which leads to a six wave mixing process (6WM). The model describes the enhancement of the amplification and intensity difference squeezing due to the interaction of the dressing field, according to the experimental parameters.
Moreover, the model predicts that this mechanism allows the spectral control of the quantum correlations between pair of side-bands by just tuning accordingly the dressing field. 
\end{abstract}

%%%%%%%%%%%%%%%%%%%%%%%%%%  body  %%%%%%%%%%%%%%%%%%%%%%%%%%
\section{Introduction}
Four wave mixing (4WM) in hot vapors is a %has raised as 
practical system to efficiently produce pairs of quantum correlated light beams~\cite{PDLett07}. More specifically, in the frequency domain, 4WM produces entanglement between two pairs of side-bands~\cite{PDLett09}. The parametric amplification (PA) due to the $\chi^{(3)}$ non-linearity leads to the squeezing of the intensity difference which is produced in a narrow bandwidth around the two photon detuning between the pump and probe. 
Such quantum properties are now exploited to generate multi-mode states\cite{Jing18} and more recently the possibility of implementing an optical parameter oscillator (OPO)~\cite{Alvaro20}  operating directly at the atomic wavelength.

In order to obtain the optimum quantum correlations in 4WM, the probe field has to be carefully tuned within the amplification bandwidth. Whilst the phenomenological model~\cite{PDLett08} indicates that looking for maximum gain in the parametric amplification leads to maximum squeezing for noise intensity difference,  experiments show that the maximum squeezing level is reached when the probe is red detuned from maximum gain two-photon resonance. This feature was later demonstrated by Quentin et al.\cite{Quentin10} with a proper microscopical model showing the high efficiency of atomic vapors to reach high levels of squeezing. In fact, 
in order to further increase the squeezing of the intensity difference, cascaded PA  4WM based processes have been proposed~\cite{Jing14}. Pursuing high levels of quantum correlations are of great importance, for instance, to reach fault-tolerant threshold  for continuous variable cluster state (CVCS) for one way quantum processing~\cite{Pfister20}.

Recently, Zhang et al. in ref.~\cite{Zhang17}  have presented an alternative method. They have proposed the use of dressing fields to further enhance the parametric gain as well as the noise intensity squeezing. An additional counterpropagating field that couples the excited level to the 5D level (of rubidium atoms), can dress the atomic level  generating a superposition of a 4WM process with a Six Wave mixing~(6WM) process. This superposition enhances the amplification gain and the noise intensity squeezing. This process has been only  described by a semi-classical model\cite{Zhang18}
% in which the optical non linearities $\chi^{3}$ and $\chi^{5}$ are calculated semicla
considering superposition of $\chi^{3}$ and $\chi^{5}$ with prediction of the maximum gain attainable. However, the semi-classical approach applied to the phenomenological model, can not be employed to obtain  predictions of gain profile and noise intensity difference spectrum with respect the light-atom interaction parameters, like Rabi frequency, one-photon and two-photon detuning, propagation length, among others.
It is an attractive possibility to have extra access to parameters allowing the fine control of the quantum correlations in entangled fields, adjusting the noise compression level and its spectra accordingly to the desired goals involving applications ranging from metrology to quantum information.
%Nevertheless, the possibility of enhancing, controlling or modulating the quantum correlations of the 4WM system by coupling an extra light field, like tweezers tweaking small parts of a big machine, is not possible in  
These controls are generally absent in solid state and chip sources that generate twin beams~\cite{Ou92,Villar05,Avik15}.

In this paper we present a microscopic description of dressed 4WM which goes beyond the phenomenological approach. Following the model presented in ref~\cite{Quentin10}, this work focus on a theoretical description of dressed 4WM mixing through a 6WM based on Heisenberg-Langevin equations on the double-$\Lambda$  system in a thick medium interacting with pump, probe and conjugate modes, and with the dressing field. We obtain a complete description of the dressed 4WM that takes into account all the interactions parameters, one-photon and two-photon detuning, Rabi frequencies, atomic density, propagation length, among others. Our model describes very closely the gain profile modified by the dressing field  and the noise intensity spectrum measured in ref.~\cite{Zhang17}. In addition, we show how the dressing field can tune the quantum correlation between the pair of side-bands, which could be detected %only 
by auto-homodyne detection~\cite{Villar08}.

\section{Heisenberg-Langevin equations for Six Wave Mixing: Dressing the four wave mixing}

\begin{figure}[h!]
\centering
\begin{overpic}[width=12cm]{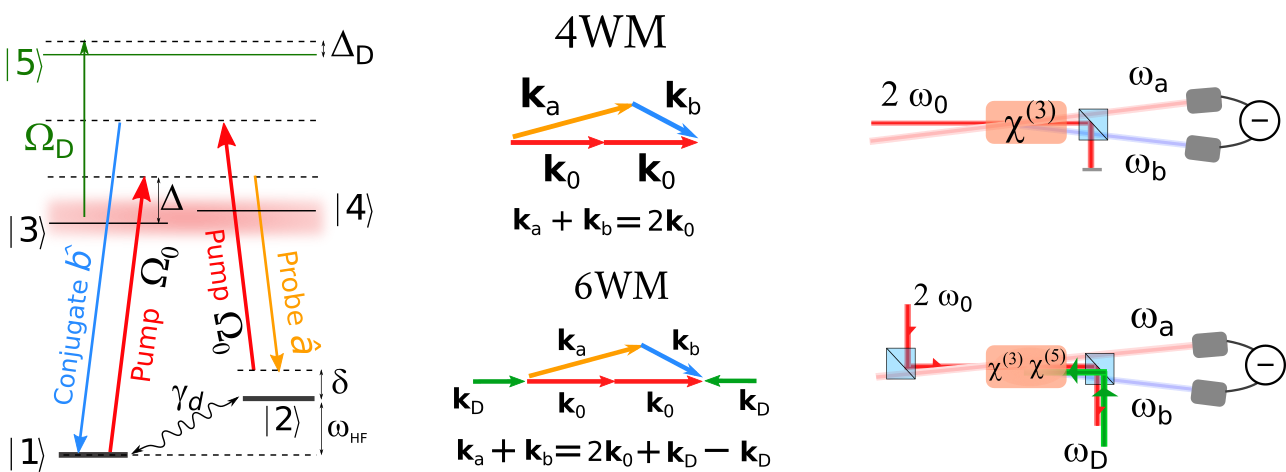}
\put(-5,30){(a)}
\put(33,30){(b)}
\put(60,30){(c)}
\end{overpic}
\caption{(a) Energy level diagram with the double $\Lambda$ scheme on the D1 line for $^{85}$Rb. 
%The hyperfine splitting is not resolve due to Doppler broadening.
(b) Geometric configuration for 4WM and 6WM. (c)
Setup to obtain 4WM and 6WM.}
\label{fig:Setup}
\end{figure}

We consider an ensemble of $N$ atoms within a cylinder  with cross section $A$ and length $L$,  having a double $\Lambda$ coupling of four levels as in Fig.\ref{fig:Setup}(a), with two ground states $|1\rangle$ and $|2\rangle$ and two excited states $|3\rangle$ and $|4\rangle$. The two excited states are subjected to spontaneous emission rate $\Gamma$ whereas the two ground state coherence can decay at a rate $\gamma$.  In the case of rubidium 87, at high temperature, the excited states are Doppler broadening, which suits for the description of this double lambda configuration. The pump beam propagates in the $\mathbf{k}_0$ direction and the probe beam is slightly shifted from the $\mathbf{k}_0$ direction propagating towards
%$\mathbf{k}_a$
$\mathbf{k}_a$, as in Fig.\ref{fig:Setup}(b). The interaction of the pump and a probe field with frequencies $\omega_0$ and $\omega_a$
drives the $\Lambda$ subsystem $|1\rangle\rightarrow |3\rangle\rightarrow |2\rangle$, whereas the interaction of the pump and the conjugate with frequencies $\omega_0$ and $\omega_b$ drives the second $\Lambda$ subsystem $|2\rangle\rightarrow |4\rangle\rightarrow |1\rangle$.
The 4WM process annihilates two photons from the pump to generate a pair involving an additional probe photon and a conjugate photon satisfying the phase matching condition $2\mathbf{k}_0=\mathbf{k}_a+\mathbf{k}_b$ and energy conservation $2\omega_0=\omega_a+\omega_b$. In general, we consider the interaction of the pump detuned by $\Delta$ with respect to the $|1\rangle\rightarrow |3\rangle$ transition, while the pump and the probe drive a Raman transition detuned by $\delta$. For the second $\Lambda$, the pump is detuned $\delta+\omega_{HF}$ and the conjugate by $\Delta+\omega_{HF}+\delta$.

%Precisa compatibilizar os campos:  probe,conjugate ou a,b? N'veis 3 e 4 precisam ser separados na figura, acompanhando as contas e a imagem construída pela tese deo Quentin (Fig. 4.5 da tese).

% \HM{Prova e conjugado nas figuras, unificar}

In addition to the three
%four 
fields used in the standard 4WM process, we add the dressing field from level $|3\rangle\rightarrow|5\rangle$ counter propagating with respect to the pump direction. Hence, we are extending the model that describes the  intensity correlations between the probe and the conjugate beam on a standard 4WM to a configuration of 6WM, as in Fig\ref{fig:Setup}(a). 

The atomic Hamiltonian is $\hat{H}_{atom}=\int_0^L dz A \rho E_{i}\hat{\sigma}_{ii}(z)$,  where $\rho$ is the atomic density, $A$ is the cross-section of interaction and $L$ is the length of propagation. Meanwhile, 
the interaction Hamiltonian with the two  pump beams and the two quantum fields $\hat{a}$ and $\hat{b}$ is given by
\begin{align}
 \hat{H}_{int}=&\hbar \int_{0}^L\left(%Destroing photons for a_+ ------------------------------------------------------
 \Omega_0 e^{i(\omega_0t-\mathbf{k}_0\cdot\mathbf{r})}\hat{\sigma}_{31}+\Omega_0\hat{\sigma}_{42}e^{i(\omega_0t-\mathbf{k}_0\cdot\mathbf{r})}+\Omega_D\hat{\sigma}_{35}e^{i(\omega_Dt-\mathbf{k}_D\cdot\mathbf{r})}\right.\nonumber\\
 &\left.+g_0^a\hat{a}(z,t)\hat{\sigma}_{32}e^{i\omega_0^at}+g_0^b\hat{b}(z,t)\hat{\sigma}_{41}e^{i\omega_0^bt} + \text{h.c.}\right)\rho A dz,
 \end{align}
where $\Omega_0=\mathbf{d }\cdot\mathcal{E}_0/\hbar$ is the dominant pump Rabi frequency and $g_0^a=p_{31}e_a$ and $g_0^a=p_{42}e_b$ are the atom-field coupling constants with $e_{a,b}=\sqrt{\hbar \omega_0^{a,b}/2\epsilon_0V}$ in which $V=A\ L$ is the quantization volume. The transition dipole moments are defined as $p_i$. These are the coupling parameters of the standard 4WM. In addition to this, $\Omega_D$ and $\omega_D$ represent the Rabi frequency and the optical frequency of the dressing field, respectively, which is consider a classical field.

As described in ref.~\cite{Quentin10} the evolution of the atomic operator is described by the Heisenberg-Langevin equations $\partial_t \hat{\sigma}_{ij}=-(i/\hbar)[\hat{\sigma}_{ij},\hat{H}_{int}+\hat{H}^1] -\sum_{n,m}\Gamma_{nm}\hat{\sigma}_{nm}+\mathcal{\hat{F}}_{ij}$, where $\Gamma_{nm}$ describe spontaneous emission and decoherence rates, induced by the stochastic operators  $\mathcal{\hat{F}}_{ij}$ which satisfy $\langle\mathcal{\hat{F}}_{ij} \rangle=0$ and $\langle\mathcal{\hat{F}}_{ij}(z,t) \mathcal{\hat{F}}_{nm}(z',t')\rangle=\delta_{ij,nm}\delta(z-z')\delta(t-t')$.
The stochastic operators describe the interaction of atoms with vacuum modes.
% Following the same procedure 
% Now we need to choose on which rotating frame we are going to solve the dynamics.
Now, choosing $\Omega_i e^{i(\omega_i t-\mathbf{k}_i\cdot\mathbf{r})}$ with $i=0$ and $D$, as our reference frame, we obtain the following dynamics in the Liouville space
%  \begin{align}
% \frac{d\mathbf{\hat{X}}(z,t)}{dt}=&[\mathbf{M}^{(0)}+\sum_{n=1}^\infty\mathbf{M}^{(n)}e^{in\Delta\mathbf{k}\cdot\mathbf{r}} +\mathbf{M}^{(-n)} e^{-in\Delta\mathbf{k}\cdot\mathbf{r}}]\mathbf{\hat{X}}(z,t)\nonumber\\
% &+ \mathbf{G}_{x} \mathbf{\hat{A}}(z,t)+\mathcal{\hat{F}}(z,t),\label{eq:DynTwoPumps_z}
% \end{align}
 \begin{align}
\frac{d\mathbf{\hat{X}}(z,t)}{dt}=&\mathbf{M}
\mathbf{\hat{X}}(z,t)+ \mathbf{G}_{x} \mathbf{\hat{A}}(z,t)+\mathcal{\hat{F}}(z,t),\label{eq:Dyn5Level}
\end{align}
where we have defined the atomic and stochastic vectors $\mathbf{\hat{X}}(z,t)=[\hat{\sigma}_{11},\hat{\sigma}_{12},\cdots,\hat{\sigma}_{55}]$ and $\hat{\mathcal{F}}(z,t)=[\hat{\mathcal{F}}_{11},\hat{\mathcal{F}}_{12},\cdots,\hat{\mathcal{F}}_{55}]$, respectively, whereas the light vector is defined as $\hat{\mathbf{A}}(z,t)=[\hat{a}(z,t),\hat{a}^\dagger(z,t), \hat{b}(z,t),\hat{b}^\dagger(z,t)]$. The dynamics is driven by the matrix $\mathbf{M}$ which describes  the single-pump matrix generator containing the main parameters of the interaction. The matrix $\mathbf{G}_{x}$ depends on the atom-field coupling constants and the mean value of the atomic operators when atoms are only interacting with the pumps along $\mathbf{k}_0$.

The propagation of the probe  and conjugate fields $\hat{a}(z,t)$ and $\hat{b}(z,t)$ through the medium are given by 
Heisenberg equation $(\partial_t + c\partial_z) \hat{O}=-(i/\hbar)[\hat{O},\hat{H}_{int}]$, 
% Defining $\hat{\mathbf{A}}(z,t)=[\hat{a}(z,t),\hat{a}^\dagger(z,t),\hat{b}(z,t),\hat{b}^\dagger(z,t)]$ the propagation of the fields is given by 
such that the propagation of the fields is 
\begin{align}
(\partial_t + c\partial_z)\hat{\mathbf{A}}(z,t)=N\mathbf{T}\mathbf{\hat{X}}(z,t),\label{eq:Aprop}
\end{align}
where $N$ is the number of atoms and the matrix $\mathbf{T}$ is a $4\times16$ elements directly proportional to the atom-light coupling constants $g_0^a$ and $g_0^b$.
% is modified by the extra phase $e^{\Delta k_z z}$.

% Since the dynamics of the atomic operator is modified by the extra phase $e^{i\Delta k_z z}$, its steady state is not trivial, applying the same method as in ref(). The extra phase affect the dynamics of the atomic operators as well as the propagation of the fields.

\section{Parametric gain of the optical modes}

The intensity gain of the modes is determined neglecting the stochastic term, %i.e last term in the r.h.s. of eq.(\ref{eq:XHarm}).
and considering the steady state condition $\partial_t \mathbf{X}=0$ and $\partial_t \mathbf{A}=0$, such that $
\mathbf{\hat{X}}^s=\mathbf{M}^{-1}\mathbf{G}_x \mathbf{\hat{A}}(z,t) $. Substituting in eq.(\ref{eq:Aprop}), the amplitude of the fields at the output of the propagation is 
% \begin{align}
% \mathbb{\hat{A}}(z)&=e^{\mathbb{R}\ z} \mathbb{\hat{A}}(0)=\mathbb{J}(z)\mathbb{\hat{A}}(0)\label{eq:SolAHarm_z}
% \end{align}
\begin{align}
\mathbf{\hat{A}}(z)&=e^{\mathbf{R}z}\mathbf{\hat{A}}(0),\label{eq:SolA_z}
\end{align}
 defining 
$\mathbf{R}=N\ \mathbf{T}\mathbf{M}^{-1}\mathbf{G}_x/c$ as the generator of the propagation.
Is worth to mention that this equation, which is considering the interaction of 5 fields, takes the same form as the 4WM case that is usually employed. Thus, 
the matrix representation suppress the algebraic complexity of the addition of the dressing field. By coupling the dressing field to the excited level, the elements of the generator $\mathbf{R}$ changes according to the  Rabi frequency and optical detunings, but not its dimensions.

In order to obtain the gain of each mode, we construct the gain matrix $\mathbf{C}(z)=\langle\mathbf{\hat{A}}(z),\mathbf{\hat{A}}(z)^T\rangle$ which in terms of the input mode fields is expressed as
\begin{align}
\mathbf{C}(z)&=\mathbf{J}(z)~\langle\mathbf{\hat{A}}(0),\mathbf{\hat{A}}(0)^T\rangle~\mathbf{J}(z)^T,
\end{align}
 where we have defined the propagation matrix  $\mathbf{J}(z)=e^{\mathbf{R}z}$ and the input matrix as $\langle\mathbf{\hat{A}}(0),\mathbf{\hat{A}}(0)^T\rangle$.
The gain intensity for each mode is  given by $G_{(a,b)}=\langle \mathbf{\hat{A}}(z),\mathbf{\hat{A}}^T(z)\rangle_{(a,b)}/|\alpha|^2$, considering the input state as a coherent state
\begin{align}
\langle\mathbf{\hat{A}}(0),\mathbf{\hat{A}}(0)^T\rangle&=\begin{bmatrix}
   \alpha^2 & |\alpha|^2 +1  & 0 & 0\\
    |\alpha|^2 & \alpha*^2 & 0 & 0\\
    0 & 0 & 0 & 1\\
     0 & 0 & 0 & 0
\end{bmatrix},
\end{align}
in which $|\alpha|^2 +1\approx|\alpha|^2$ for intense light beams.

\begin{figure}[h!]
\centering
\begin{overpic}[width=11cm]{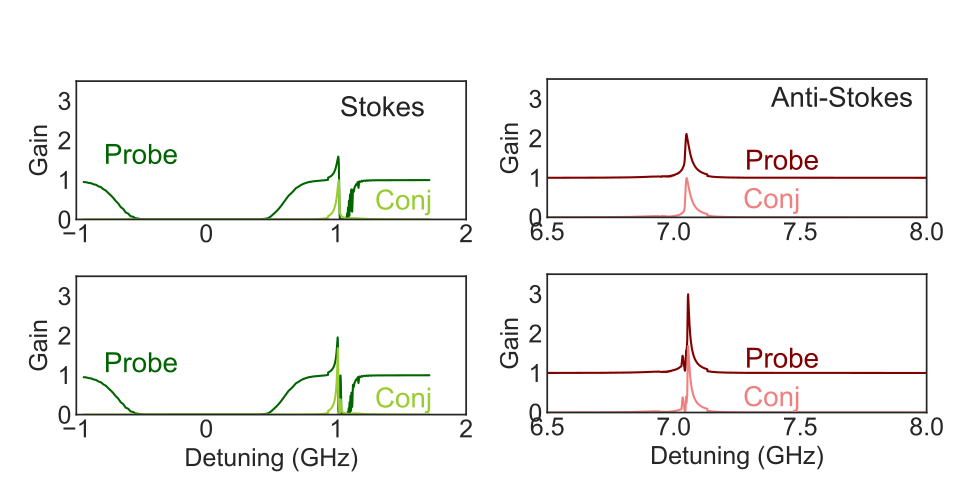}
\put(-3,40){(a)}
\put(-3,20){(b)}
\end{overpic}
% \begin{overpic}[width=3cm]{Gain_spectrum_As.png}
% \end{overpic}
% \includegraphics[width=8cm]{Gain_spectrum}\quad
% \includegraphics[width=3cm]{Gain_spectrum_As}
% \begin{overpic}[width=18cm]{Gain_spectrum.pdf}
% \put(0,60){(a)}
% % \put(0,20){(b)}
% % \put(45,60){(c)}
% \end{overpic}
\caption{Amplification spectrum of the probe and conjugate beam for the stokes and anti-stokes channels of 4WM, integrated over the Doppler broadening. (a) 4WM without dressing field. (b) 4WM with the dressing field. The parameters of the calculation are: $\Omega_0/2\pi=480$~MHz, $\Gamma/2\pi=5.7$~MHz, $\gamma_D/2\pi=1$~MHz, $\omega_{HF}/2\pi=3.035$~GHz, $\Omega_D=0.2~\Omega_0$,  and $\Delta_D=-1.04$~GHz. }
\label{fig:4WM_dressed}
\end{figure}

From this solution one can obtain the gain spectroscopy for the 4WM process in the presence of the dressing field. Figure \ref{fig:4WM_dressed} shows the typical gain
(defined as the ratio of each output power to the input probe field) %amplification gain spectrum
as a function of the frequency of the probe beam close to the D1 line integrated over the group of velocities comparing the effect of the dressing field. In particular, Figure (a) shows the spectrum for the standard 4WM for the Stokes and anti-Stokes channel when no dressing field is used.
In such situation the anti-Stokes channel, for instance, reaches a maximum  gain factor of $G\sim2$. On the other hand, figure (b) shows the gain spectrum when the dressing field interacts with the atoms detuned $\Delta_D=-1$GHz from the excited state. Notice that the amplification gain is  increased  by an extra unit % factor of $\sim1$ 
for both beams, consistent with the observation in ref.~\cite{Zhang17}. 
% \subsection{Fields Fluctuations}
% Effect of the extra pump M_3
% \onecolumngrid
%Colocar as figuras: à esquerda, só o canal Stokes, à direita, só o detalhe no canal Anti-Stokes, eliminando a redundância.

%Tem alguma explicaçào física para este processo?

The model allows us to investigate the tunability of the 6WM process with respect to different parameters of the atom-light interaction, as Rabi frequency, one and two photon detunig and atomic density. In particular, this is done for the anti-Stokes channel which we can compare with the experimental observations.

\subsection{Gain optimization with respect to dressing field Rabi frequency}
\begin{figure}[h!]
\centering
\begin{overpic}[width=13cm]{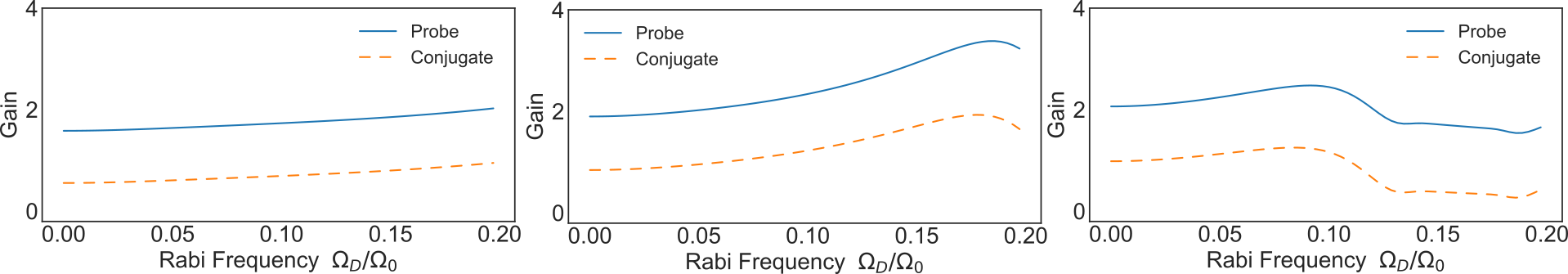}
\put(3,15){(a)}
\put(60,5){(b)}
\put(70,15){(c)}
\end{overpic}
\caption{Gain of amplification as a function of the Rabi frequency of the dressing field normalized by the Rabi freuqnecy of the pump beam. Dressing detuning (a) $\Delta_D=-1.002$~GHz (b) $\Delta_D=-1.012$~GHz. (c) $\Delta_D=-1.017$~GHz. All the other parameters are the same as in Fig.~\ref{fig:4WM_dressed}}
\label{fig:gain_vs_g5}
\end{figure}

First, we explore the gain amplification with respect to intensity of the dressing field.
For an specific detuning of the dressing field, its intensity can adjust the maximum amplification. Figure \ref{fig:gain_vs_g5} shows the amplification gain of the anti-stokes channel as a function of the Rabi frequency of the dressing field, for three different optical detunings $\Delta_D$.
We have three different scenarios of amplification within a range of 20\% ratio between the dressing field and the pump field.
Figure (a) shows the first case with $\Delta_D=-1.002$~GHz, in which  the Rabi frequency of the dressing field increases the gain amplification monotonically. Meanwhile,  Figure (b) shows the gain for $\Delta_D=-1.012$~GHz closer to the two-photon detuning of the 4WM, in which the gain amplification reaches its maximum at $\Omega_D\sim0.19\Omega_0$. Finally, Figure (c) for at $\Delta_D=-1.017$~GHz, shows the frequency shift due the AC stark shift, reaching its maximum at $\Omega_D\sim0.1\Omega_0$ but in this case with a factor of 0.6 improvement of gain.
Within the range of values reported in ref.~\cite{Zhang17}, the case in figure(b) seems to be consistent with the experimental observations.

It is worth noticing that higher gain can be obtain with larger Rabi frequency. However, an interesting situation is when the increase of amplification is obtained with the lowest extra energy possible, instead of converting a co-propagating 4WM into a counter-propagating 4WM. For weak Rabi frequencies, the superposition of the 4WM and the 6WM due by dressing the atomic states is what boost the amplification process. For high Rabi frequencies, the system runs in a 4WM with two counter-propagating pumps.

\subsection{Gain control with dressing field detuning}

\begin{figure}[h!]
\centering
\begin{overpic}[width=6.2cm]{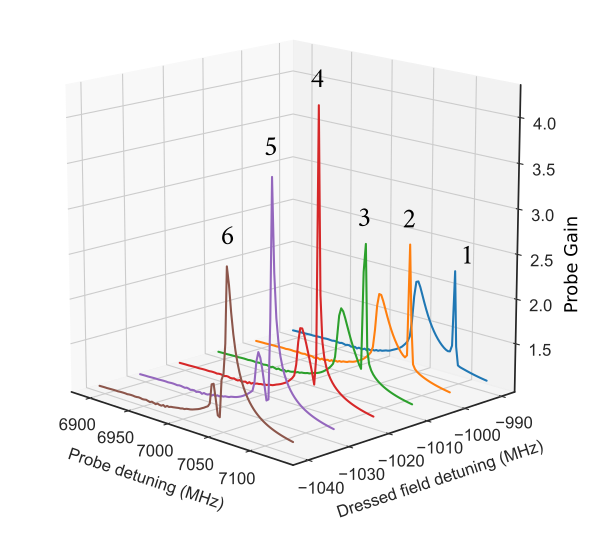}
\put(0,70){(a)}
\end{overpic}
\begin{overpic}[width=6.2cm]{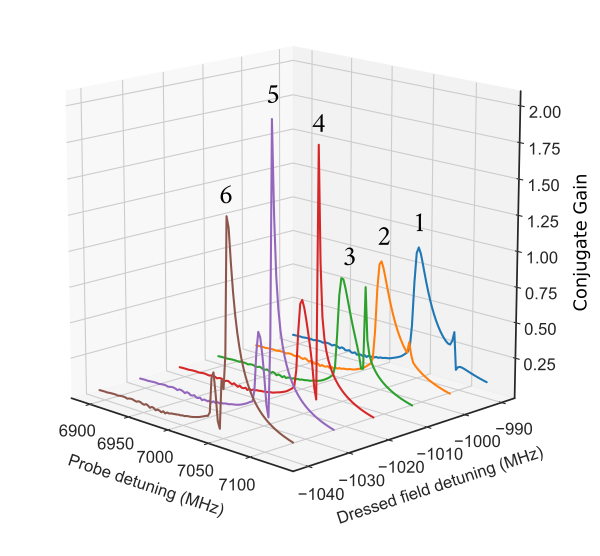}
\put(0,70){(b)}
\end{overpic}
\caption{Anti-Stokes gain spectrum for different optical detunings of the dressing field $\Delta_D$. (a) Probe and (b) Conjugate. The Rabi frequency of the dressing field is $\Omega_D=0.2~\Omega_0$. All the other parameters are the same of Fig.~\ref{fig:4WM_dressed}}
\label{fig:gain_vs_delta5}
\end{figure}

We also investigated the 6WM gain  with respect to the dressing field detuning for a fixed Rabi frequency. Figure \ref{fig:gain_vs_delta5} shows the probe and conjugate gain spectrum for different optical detunings of the dressing field. The curve 1 shows the narrow dressed 4WM resonance detuned from the maximum of the 4WM gain, when the dressing field detuning is $\Delta_D=-990$~MHz. In this situation the dressing field just modifies the $\chi^{(3)}$ response leading to a dispersive profile.
By increasing the dressing field detuning, the 6WM starts to become generated and superposed to the 4WM, reaching the maximum amplification. Notice that in the range of $\Delta_D\sim-1.020-1.030$~GHz (curves 4 and 5), the increase of amplification is followed by steep dispersive profile, whereas for  $\Delta_D\sim-1.040$~GHz the increase of amplification is added by a factor of 1 where the dispersive profile is not as steep as the case of $\Delta_D\sim-1.020$~GHz.

The control over the %e tunability of
amplification gain in only one vapor cell 
%is what
makes this mechanism interesting for implementations, when compared to those proposals where secondary 4WM is used as an extra amplification stage \cite{Jing14}, occupying extra space and synchronization of both systems.

\section{Noise properties of Dressed 4WM}

Next, in order to analyze the field fluctuations in the frequency domain, we employ the linearization $\hat{O}=\langle \hat{O}\rangle + \delta\hat{O}$ for the atomic and light operators. Following the same procedure as in ref.~\cite{Quentin10}, we apply the Fourier transform to the dynamical equations (\ref{eq:Dyn5Level}) and the propagation equation (\ref{eq:Aprop}). The atomic operators' fluctuations are $
\delta\mathbf{\hat{X}}(\omega)=-[i\omega\mathbf{I}+\mathbf{M}]^{-1}[ \mathbf{G}_x \delta\mathbf{\hat{A}}(\omega)+\hat{\mathcal{F}}(z,t)]$.
 Thus the optical fields modes fluctuations follow the propagation equation
\begin{align}
\frac{\partial \delta \mathbf{\hat{A}}(\omega)}{\partial z}&=\mathbf{R}(\omega)\delta \mathbf{\hat{A}}(\omega)+\mathbf{R}_F(\omega) \mathcal{\hat{F}}(z,\omega),\label{eq:d_deltaAFloqExp}
\end{align}
where $\mathbf{R}(\omega)=-(N/c)\mathbf{T}\ \tilde{\mathbf{M}}(\omega)\ \mathbf{G}_x +i(\omega/c) \mathbf{I}$ and $\mathbf{R}_F(\omega)=(N/c)\mathbf{T}\tilde{\mathbf{M}}(\omega)$, in which we have defined $\tilde{\mathbf{M}}(\omega)=[i\omega\mathbf{I}+\mathbf{M}]^{-1}$. Notice that this equation takes a similar form as the carrier solution in eq.(\ref{eq:SolA_z}), except by the additional stochastic term $\mathcal{\hat{F}}(z,\omega)$.
Hence, the solution for the propagation of the fields is given by
% $\delta\mathbb{\hat{A}}(z,\omega)=e^{\mathbb{R}(\omega)\ z} 
% \delta\mathbb{\hat{A}}(0,\omega)
% $ which can be re written as 
\begin{align}
\delta\mathbf{\hat{A}}(z,\omega)&=\mathbf{J}(z,\omega)
\delta\mathbf{\hat{A}}(0,\omega)+\mathbf{J}(z,\omega) \mathcal{\hat{F}}_{in}(z,\omega),\label{eq:SolFFTdA_n}
\end{align}
defining the propagator as $\mathbf{J}(z,\omega)=e^{\mathbf{R}(\omega)\ z}$ and the stochastic term as $\mathcal{\hat{F}}_{in}(z,\omega)=(\hat{F}_{a}(z,\omega)$ ~$,\hat{F}_{a^\dagger}(z,\omega),\hat{F}_{b}(z,\omega),\hat{F}_{b^\dagger}(z,\omega))^T$ which is calculated from
\begin{align}
\mathcal{\hat{F}}_{in}(z,\omega)=&\int_0^z dz' e^{-\mathbf{R}(\omega)\ z'}\mathbf{R}_F(\omega) \mathcal{\hat{F}}(z',\omega).
\end{align}

It is worth noting that dimension of the stochastic vector $\mathcal{\hat{F}}_{in}(z,\omega)$ does not change as the number of dressing fields increases. However, the stochastic operator $\mathcal{\hat{F}}(z',\omega)$, does change with the number of levels coupled by additional dressing fields.

From the amplitude fluctuations, the power spectral density for the amplitude field operators is given by~ 
$\mathbf{S}(z,\omega)~=(2\pi c/L)~ \langle\delta\mathbf{\hat{A}}(z,\omega),\delta\mathbf{\hat{A}}(z,\omega')^T\rangle\delta(\omega+\omega')$,
% \begin{align}
% \mathbb{S}(z,\omega)=(2\pi c/L) \Braket{\delta\mathbb{\hat{A}}(z,\omega),\delta\mathbb{\hat{A}}(z,\omega')^T}\delta(\omega+\omega')\label{eq:SolFFTS}
% \end{align}
thus, expressing the power density explicitly in terms of the solution (\ref{eq:SolFFTdA_n}) one obtains
\begin{align}
\mathbf{S}(z,\omega)=&\mathbf{J}(z,\omega)\mathbf{S}(0,\omega)\mathbf{J}(z,\omega')^T+\mathbf{J}(z,\omega)\mathbf{S}_F(z,\omega)\mathbf{J}(z,\omega')^T,
\end{align}
in which $\mathbf{S}(0,\omega)=(2\pi c/L)\langle\delta\mathbf{\hat{A}}(0,\omega),\delta\mathbf{\hat{A}}(0,\omega')^T\rangle\delta(\omega+\omega')=2\pi \mathbf{S}_A(\omega) \delta(\omega+\omega')$ and the stochastic term $\mathbf{S}_F(z,\omega)=(2c/L)\langle\mathbf{\hat{F}}_{in}(z,\omega),\mathbf{\hat{F}}_{in}(z,\omega')^T\rangle$. This matrix allows us to determine the intensity  power spectrum for the pair of fields produced by the dressed 4WM process.  To do so,  we consider the quadratures of the fields by  transforming the amplitude vector into the quadrature vector like $\delta\mathbf{\hat{Y}}(z,\omega)=\mathbf{U}(z)\delta\mathbf{\hat{A}}(z,\omega)$,
such that 
\begin{align}
\delta \mathbf{Y}(z,\omega)&=\begin{bmatrix}
  e^{i\phi_a} & e^{-i\phi_a} & 0 & 0\\
    -ie^{i\phi_a} & ie^{-i\phi_a} & 0&0\\
    0&0&e^{i\phi_b^{(n)}} & e^{-i\phi_b}\\
    0&0&-ie^{i\phi_b} & ie^{-i\phi_b}
\end{bmatrix}
\begin{bmatrix}
  \delta \hat{a}(z,\omega)\\
  \delta \hat{a}(z,\omega)^\dagger\\
    \delta \hat{b}(z,\omega)\\
  \delta \hat{b}(z,\omega)^\dagger
\end{bmatrix}=\begin{bmatrix}
  \delta \hat{X}_a(z,\omega)\\
   \delta \hat{P}_a(z,\omega)\\
    \delta \hat{X}_b(z,\omega)\\
   \delta \hat{P}_b(z,\omega)
\end{bmatrix}
\end{align}
which defines the quadratures fluctuations $\delta \hat{X}_a(z,\omega)$ and $\delta \hat{P}_a(z,\omega)$ for the probe beam and equivalently to the conjugate beam. The matrix $\mathbf{U}$ is z-dependent, since the phase $e^{i\phi_a}$  and $e^{i\phi_b}$ are calculated at a particular position $z$, from eq.(\ref{eq:SolA_z}). Therefore, the covariance matrix $\mathbf{V}=\langle\delta \mathbf{Y}(z,\omega),\delta \mathbf{Y}(z,-\omega)^T\rangle$ can be written as $\mathbf{V}(z,\omega)=\mathbf{V}_{out}(z,\omega)\times 2\pi\delta(\omega+\omega')$ where
% \begin{align}
% \mathbf{V}(z,\omega)=\left[ \mathcal{L}(z,\omega)\mathbf{S}_A(\omega)\mathcal{L}(z,-\omega)^T\right]
% \times 2\pi\delta(\omega+\omega')
% \end{align}
\begin{align}
\mathbf{V}_{out}(z,\omega)=\mathcal{L}(z,\omega)\mathbf{S}_A(\omega)\mathcal{L}(z,-\omega)^T, \label{eq:cov_matrix}
\end{align}
where we have defined the overall transformation $\mathcal{L}(z,\omega)=\mathbf{U}(z)\mathbf{J}(z,\omega)$.

\subsection{Intensity noise spectrum }
In order to have a more direct comparison with the experimental observations, we first analyze the intensity difference of the dressed 4WM, which shows the quantum properties of the pair of beams generated by the 4WM process.
% The noise intensity difference spectrum can show the quantum properties of the pair of beams generated by the 4WM process. It is the first observable that is measured in experimental realizations.
Hence, the intensity fluctuation $\delta \hat{I}(z,\omega)$  is proportional to its amplitude quadrature fluctuations $\delta \hat{X}(z,\omega)$, such that for both channels $\hat{a}$ and $\hat{b}$ are defined as 
\begin{align}
\delta \hat{I}_a(z,\omega)=|\alpha|\delta \hat{X}_a(z,\omega),\\
\delta \hat{I}_b^{(n)}(z,\omega)=|\beta|\delta \hat{X}_b(z,\omega),
\end{align}
where $|\alpha|$ and $|\beta|$ are obtained from the gain solution in eq.(\ref{eq:cov_matrix}). Therefore, the sum and the difference of intensity fluctuations are $\delta \hat{I}_{\pm}(z,\omega)=|\alpha|\delta \hat{X}_a(z,\omega)~\pm~|\beta|\delta \hat{X}_b(z,\omega)$.

\begin{figure}[h!]
\centering
\begin{overpic}[width=6cm]{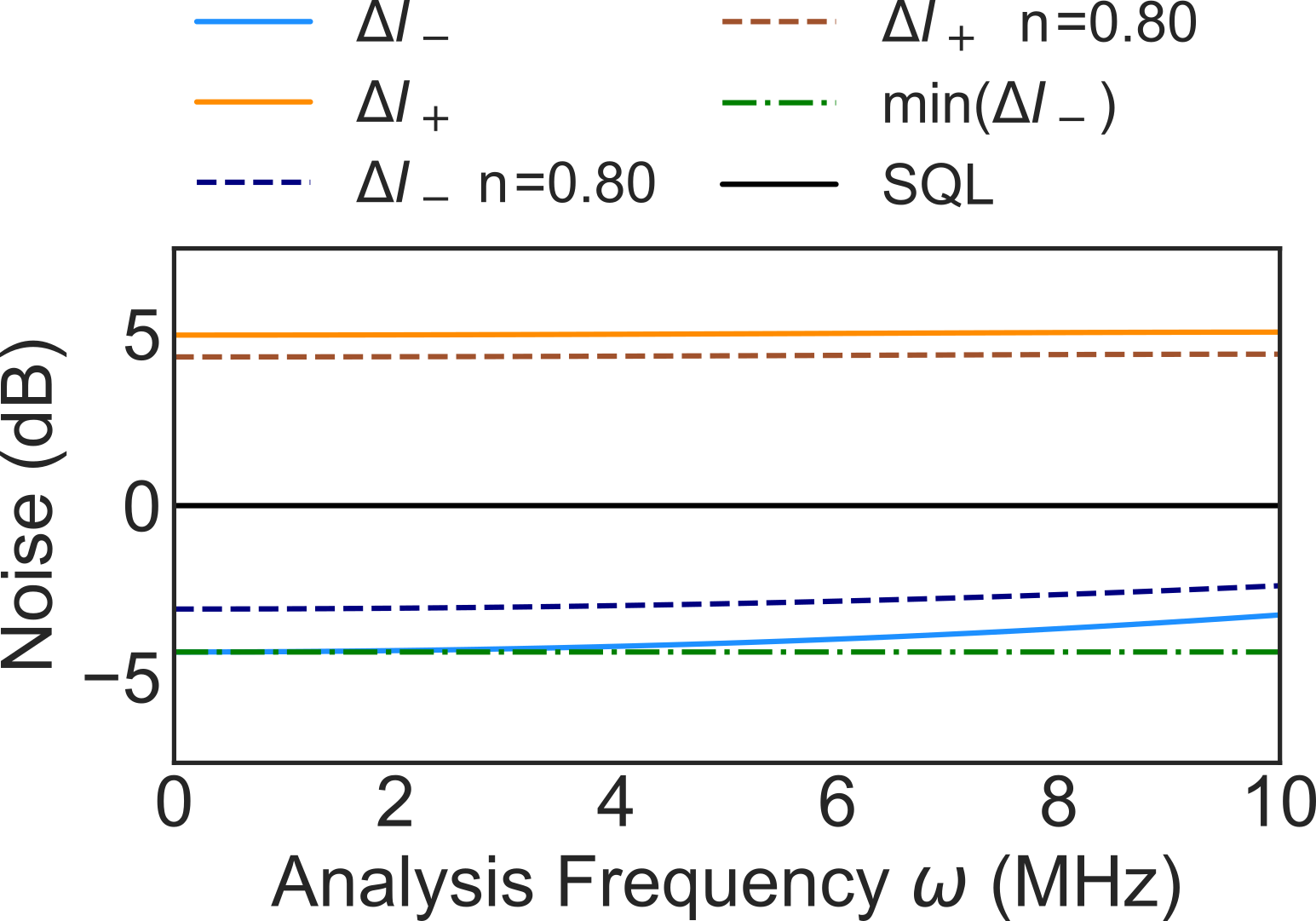}
\put(0,55){(a)}
\end{overpic}
\begin{overpic}[width=6cm]{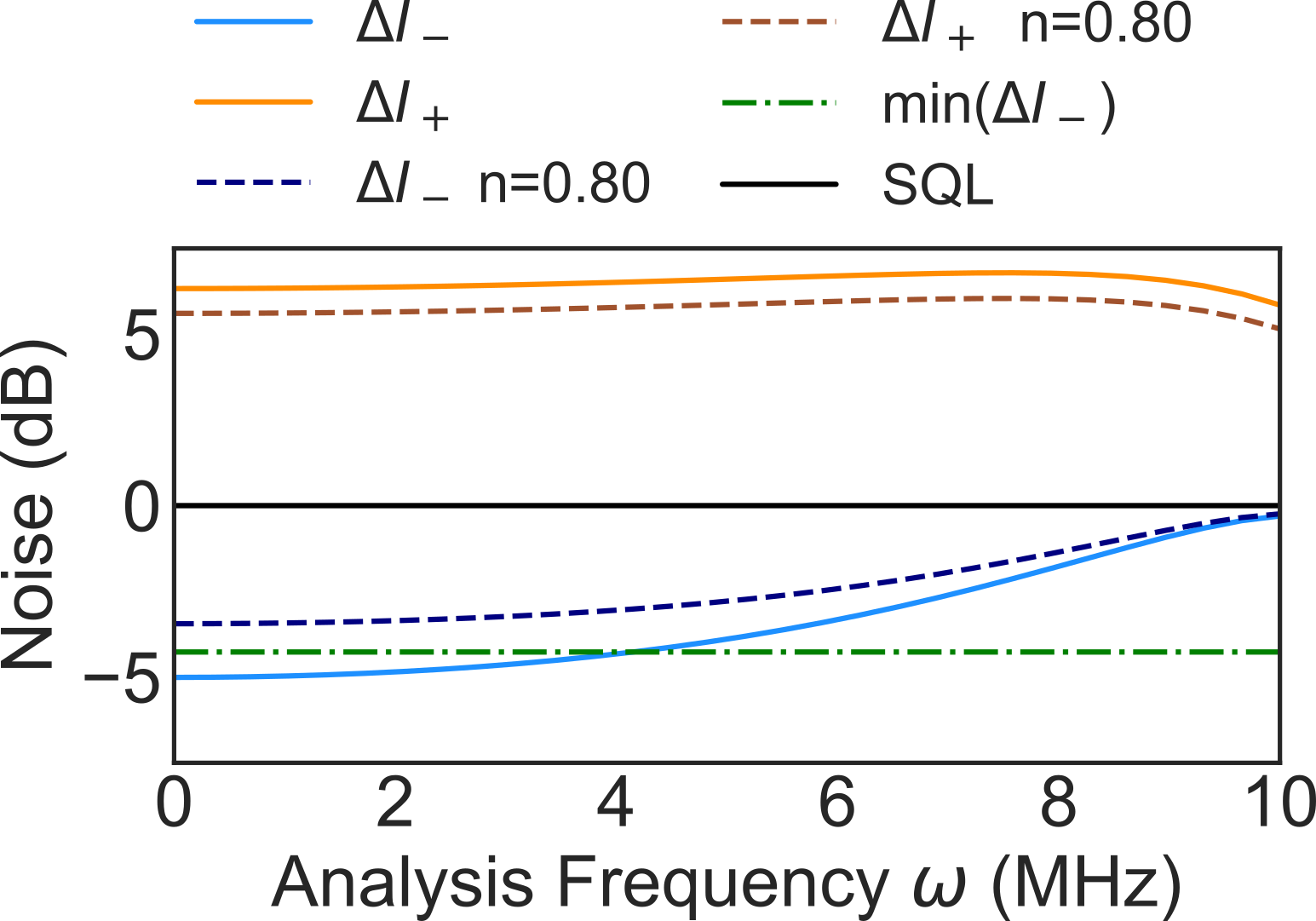}
\put(0,55){(b)}
\end{overpic}
\caption{Noise spectrum of the 4WM process (a) without dressing field and (b) with the interaction of the dressing field considering $\Omega_D=0.15\Omega_0$ and $\Delta_D=-1040$~MHz.
% \MM{confira se essa é a ideia} 
The dashed-green line corresponds to the lowest compression level $ \min(\Delta I_-)$ within the range of $\sim 2$~MHz, obtained when the dressing field is not coupled to the medium.
}
\label{fig:NoiseIntensity}
\end{figure}

Figure \ref{fig:NoiseIntensity}~(a) shows the intensity noise spectrum of the sum and the difference between the $\hat{a}$ and $\hat{b}$ modes for the typical 4WM process when the dressed field is not coupled to the atoms.
One can observe that the noise of the sum present excess of noise whereas the difference is below the shot noise within a range 10MHz. The solid line shows the squeezing for 100\% detection efficiency whereas the dashed line shows the correction by 80\% detection efficiency. Now we analyze the intensity spectrum when the dress field is coupled at $\Delta_D=-1040MHz$. Figure \ref{fig:NoiseIntensity}(b) shows that the squeezing of the intensity difference  increases for low frequencies by 1dB. This is consistent with the observations in ref.~\cite{Zhang17}. Nevertheless, this increase in the noise compression comes with a cost. One can observe the effect of the dress field on the reduction of squeezing at an analysis frequency of  10 MHz.
%for the particular detuning of the field e.g. the
It is worth noting that by changing the relative detuning of the dressing field,  the intensity noise  spectrum profile can also present a different intensity noise spectrum profile.
% On can notice that all pairs of modes present excess of noise. 

So far, the intensity difference has been the main property observed experimentally in the case of dressed 4WM. Nevertheless, the model here presented allows to extend the study to the entanglement properties of the 4WM under the interaction of the dressing field, and how those properties are controlled by the Rabi frequency and detuning of the dressing field.

\subsection{Entanglement of the dressed 4WM}
Let us now compare the effect of the dressing field over the entanglement of the probe and conjugate fields using the entanglement criterion of Duan \textit{et al.}~\cite{Duan00}. In order to test the Duan criterion, we now compute the sum and the difference of the quadratures  $\delta \hat{X}_{\pm}(z,\omega)=\delta \hat{X}_a(z,\omega)~\pm~\delta \hat{X}_b(z,\omega)$, such that if the inequality $ \Delta^2 \hat{X}_-(z,\omega)+\Delta^2 \hat{P}_+(z,\omega)\geq2$ is violated, the quadratures of the two modes are entangled.

\begin{figure}[h!]
\centering
\begin{overpic}[width=6cm]{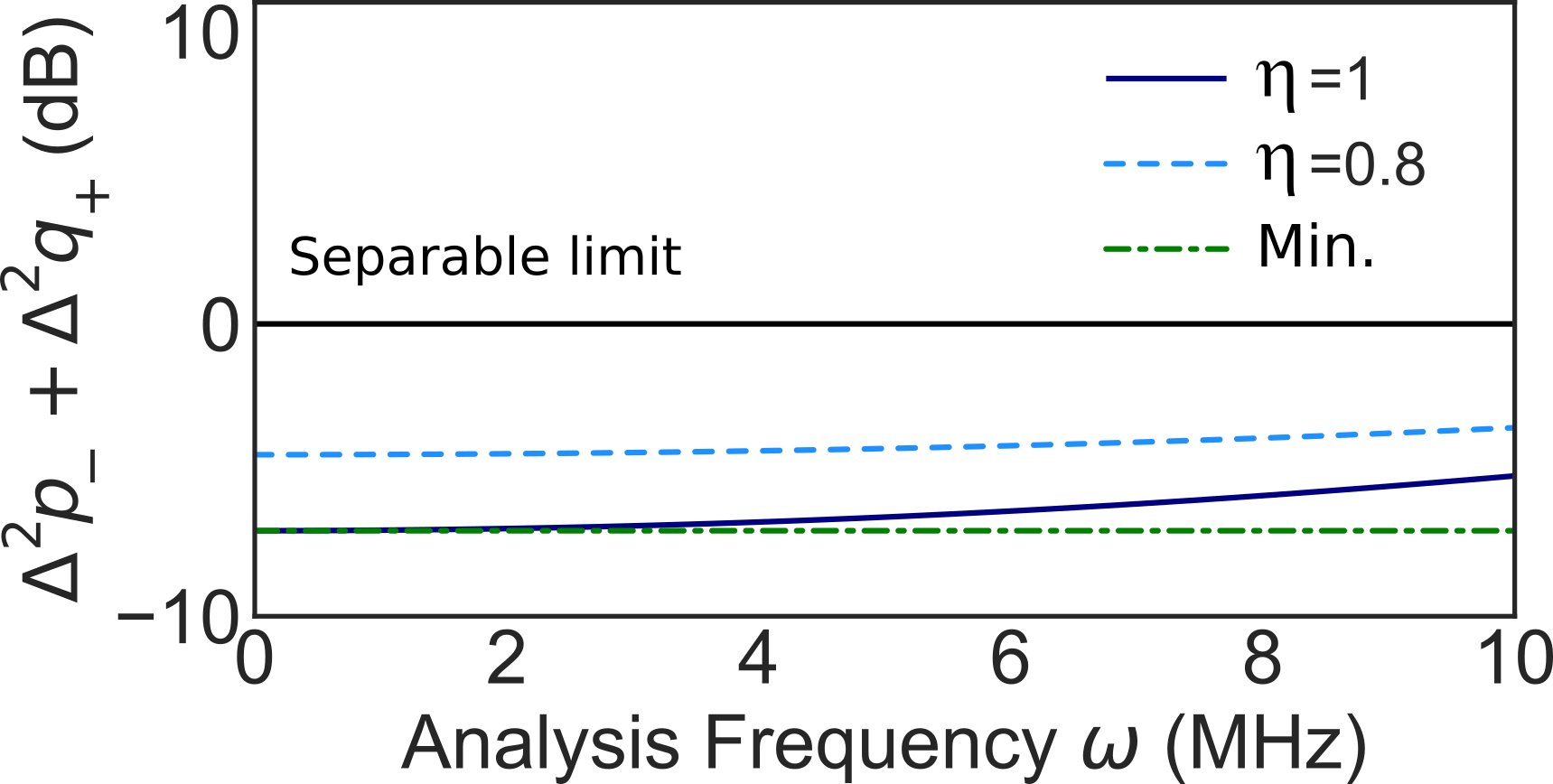}
\put(20,45){(a)}
\end{overpic}
\begin{overpic}[width=6cm]{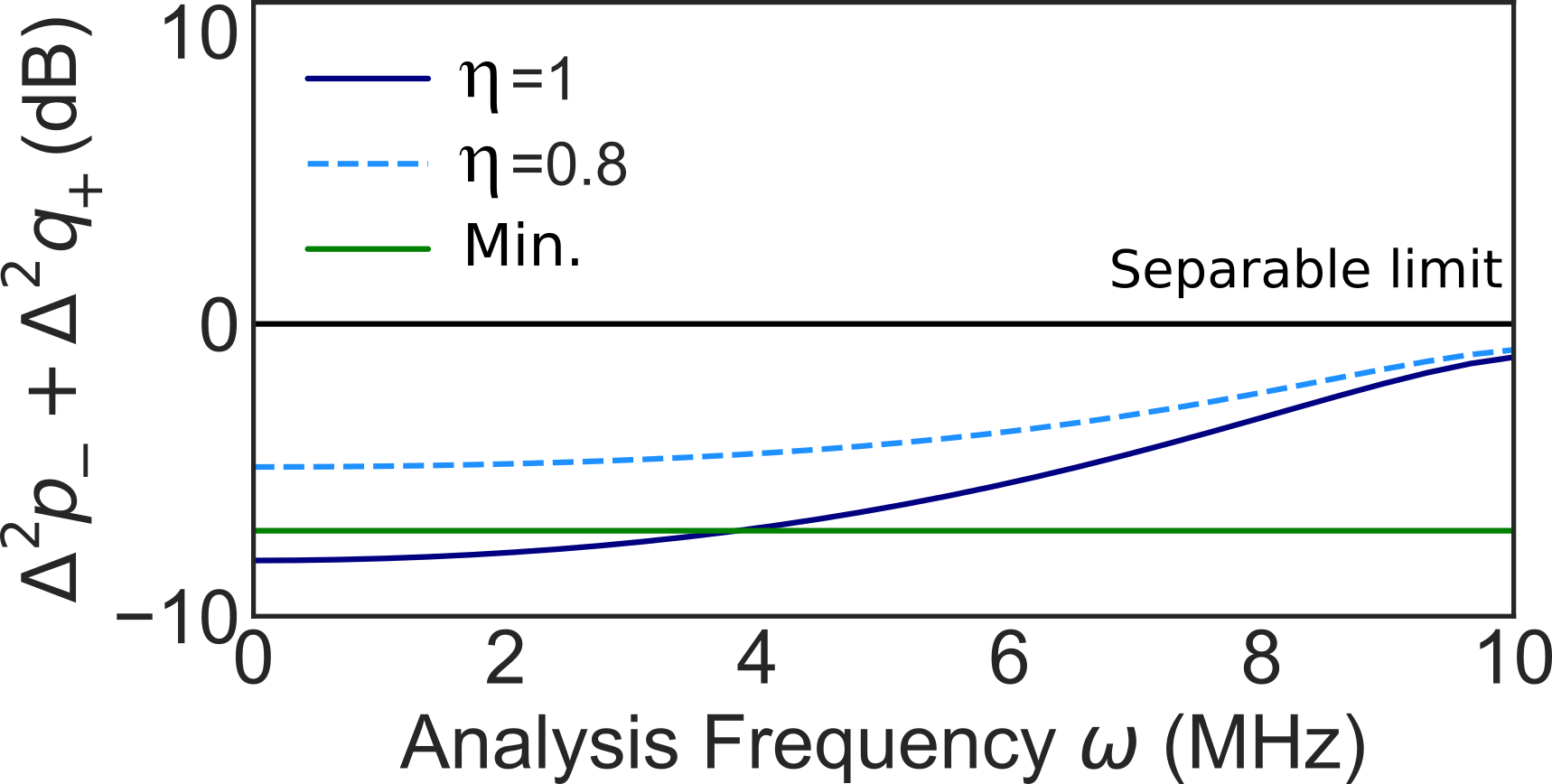}
\put(45,45){(b)}
\end{overpic}
\caption{Spectrum of the Duan criterion for separability of the probe and conjugate fields. (a) Without interaction with the dressing field. (b) With the interaction of the dressing field at $\Delta_D=-1040$~MHz. All the other parameter are the same of Fig.\ref{fig:NoiseIntensity}.}
\label{fig:dressed_entangled}
\end{figure}

Figure \ref{fig:dressed_entangled} shows the comparison of the entanglement criterion for the standard 4WM and the dressed 4WM mixing. Following the same behavior of the intensity correlation, the Duan criterion becomes more robust for lower frequencies with the dressed fields compare to the standard 4WM. Now for 10MHz of analysis frequency, the criterion tends to the classical limit. Both graphs shows the entanglement criterion with 100\% and 80\% detection efficiency.

%The effect of the dressing field on the quantum correlations between the two modes can be adjusted by its frequency.
%of the dressing field. 
In order to have a better picture of the spectral control of the quantum correlations due to the dressing field detuning, we can analyze the sidebands of the probe and conjugate independently. In what follows we analyzed the quantum correlations of the two output beams in terms of four modes of side-band frequencies, associated with the observation of the analysis frequency,  and studied the effect of the dressed field on them.

\subsection{Tuning the quantum correlation of four modes}

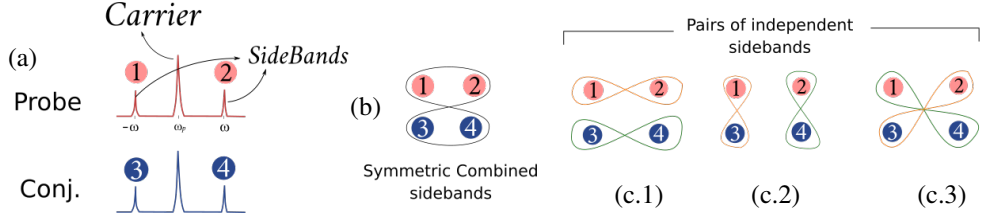
\begin{figure}[t!]
\centering
\begin{overpic}[width=13cm]{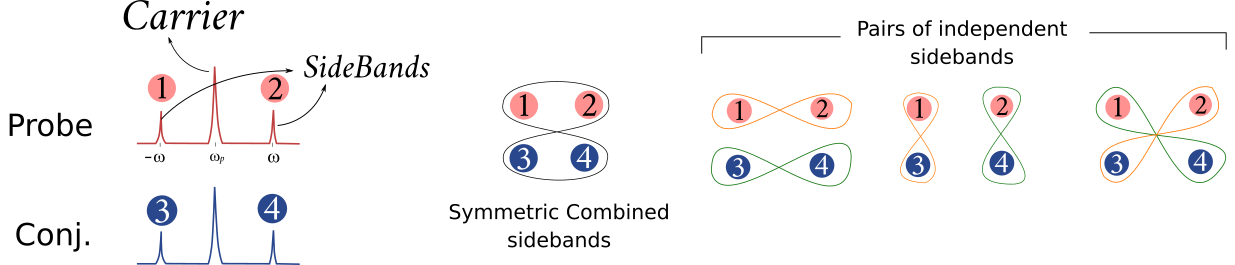}
\put(0,15){(a)}
\put(35,10){(b)}
\put(62,1){(c.1)}
\put(76,1){(c.2)}
\put(93,1){(c.3)}
\end{overpic}
\caption{(a) Representation of side-band frequencies of the light fields. (c) Representation of correlations with the bi-partition of symmetric combined side-bands for the probe and conjugate fields. (c) Representation of bi-partitions of pairs of independents side-bands.}
\label{fig:sidebands}
\end{figure}

% This two mode squeezing for the intensity difference can be analyze in terms of  the 4 modes of side-bands instead of the two pair of modes as it is done for standard measurements. 
According to refs.~\cite{Felippe13,Felippe18}, the photocurrent measured by the detectors contains information of the symmetric and anti-symmetric composition of side-bands. Hence, the quantum correlation between the sidebands of the probe and conjugate (see Fig.~\ref{fig:sidebands}(a)), are indeed correlation of the bipartition as shown in Fig.\ref{fig:sidebands}(b), and those are  typically obtained in homodyne detection. This kind of representation is associated to the fact that the quadratures are generally expressed as combinations of upper and lower sidebands, e.g.$X_\pm(\omega)=\hat{a}_\omega+\hat{a}^\dagger_{-\omega}\pm \hat{b}_\omega+\hat{b}^\dagger_{-\omega}$.
In other words, symmetric and anti-symmetric basis contains simultaneous information of upper and lower sidebands.
However, such description can be in fact decomposed into particular sidebands of the field, leading to a detailed four mode description of the field state.
%by the upper and lower sideband of frequencies of the light fields independently. 

In this case we are interested in the independent quadratures of the upper and lower sidebands such that $X_\omega=\hat{a}_\omega+\hat{a}^\dagger_\omega$ with $[\hat{a}_\omega,\hat{a}^\dagger_{\omega'}]=\delta(\omega-\omega')$.
Following ref. \cite{Felippe13}, one can transform the symmetric and anti-symmetric description of the quadrature in terms of independent side-bands transforming $\delta \hat{\mathbf{Y}}(\omega)=(\delta \hat{X}_a(\omega),\delta \hat{P}_a(\omega),\delta \hat{X}_b(\omega),\delta \hat{P}_b(\omega))^T$ into actual Hermitian operators $\delta \hat{\mathbf{Y}}_\omega=(\delta \hat{X}^a_\omega,\delta \hat{P}^a_\omega,\delta \hat{X}^b_\omega,\delta \hat{P}^b_\omega)^T$.
In order to do so, we need to double our linear vector space by defining
% \begin{align}
% \delta\hat{\mathcal{Y}}(\omega)&=(\delta \hat{\mathbf{Y}}(\omega),\delta \hat{\mathbf{Y}}(-\omega))^T\\
% \delta\hat{\mathcal{Y}}_\omega&=(\delta \hat{\mathbf{Y}}_\omega,\delta \hat{\mathbf{Y}}_{-\omega})^T
% \end{align}
$\delta\hat{\mathcal{Y}}(\omega)=(\delta \hat{\mathbf{Y}}(\omega),\delta \hat{\mathbf{Y}}(-\omega))^T$ as the
symmetric combined basis and $\delta\hat{\mathcal{Y}}_\omega=(\delta \hat{\mathbf{Y}}_\omega,\delta \hat{\mathbf{Y}}_{-\omega})^T$ for the independent sideband basis. Employing the unitary transformation $\mathbf{L}_{ab}$ defined in appendix \ref{appendix}, we can take one basis into the other one as
\begin{align}
\delta\hat{\mathcal{Y}}_\omega&=\mathbf{L}_{ab}~ \delta\hat{\mathcal{Y}}(\omega).
\end{align}
% where 
% \begin{small}
% \begin{align}
% \mathbf{L}_{ab}=\frac{1}{2}\begin{bmatrix}
%   1 & i &  0 & 0 & 1 & -i & 0 & 0\\
%     -i & 1 & 0 & 0 & i & 1 & 0 & 0\\
%     0 & 0 & 1 & i & 0 & 0 & 1 & -i \\
%      0 & 0 & -i & 1 & 0 & 0 & i & 1\\
%      1 & -i &  0 & 0 & 1 & i & 0 & 0\\
%     i & 1 & 0 & 0 & -i & 1 & 0 & 0\\
%     0 & 0 & 1 & -i & 0 & 0 & 1 & i \\
%      0 & 0 & i & 1 & 0 & 0 & -i & 1\\
% \end{bmatrix}
% \end{align}
% \end{small}
% we obtain the array of quadratures for both the beams $\hat{a}$ and $\hat{b}$ for the upper and lower  sidebands $\delta \hat{\mathbf{Y}}_{\pm\omega}$. 

% PUT A plot with sidebands representation and the spectrum
\begin{figure}[b!]
\centering
\begin{overpic}[width=13cm]{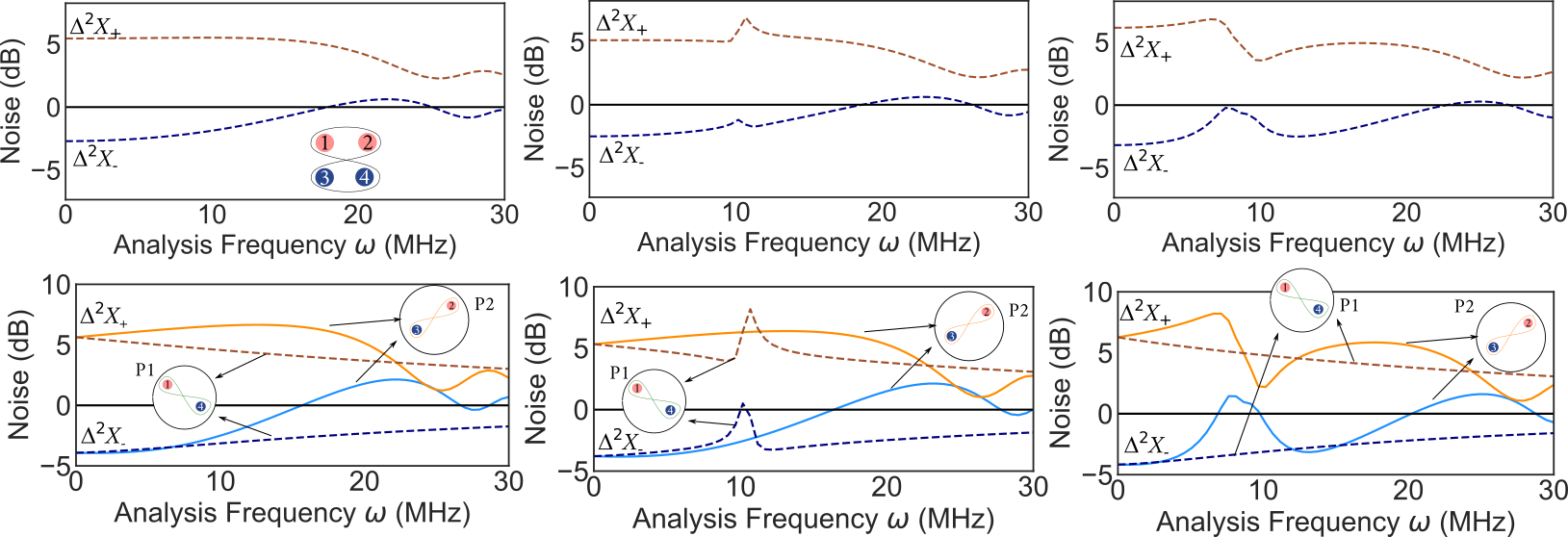}
\put(15,-4){(a)}
\put(50,-4){(b)}
\put(80,-4){(c)}
\end{overpic}
\vspace{0.5cm}
\caption{Noise spectrum of the sum and difference of amplitude quadratures for the symmetric basis (top) and independent basis (bottom). (a) 4WM without dressing field. (b) 4WM with interaction of the dressing field at $\Delta_D=-1010$~MHz and (c) at $\Delta_D=-1030$~MHz.}
\label{fig:4mode_dressing}
\end{figure}

From this transformation one has access to the quadratures of the upper and lower  sidebands $\delta \hat{\mathbf{Y}}_{\pm\omega}$ for both  beams $\hat{a}$ and $\hat{b}$ as well as the correlations of all possible bipartitions of the system.
We are particularly interested in those with non-zero correlations shown in Fig.~\ref{fig:sidebands}(c3). Another interesting feature of this treatment comes from the revelation of the multipartite nature of the generated state, doubling its space from 2 to 4 modes, leading to a detailed description that can be useful for quantum information processing and investigation of multimode entanglement.

Figure \ref{fig:4mode_dressing} compares the difference and the sum of the amplitude quadratures for different detunings of the dressing field and its role over the control of those correlations.
In this particular case the probe is red detuned $\delta=-4$MHz from the pump frequency, which is below the maximum gain of the in Fig.~\ref{fig:4WM_dressed}.
The first row shows the noise spectrum for symmetric sideband basis and the second row for two pairs of independent side-bands. Figure (a) shows the noise spectrum without the interaction of the dressing field. Notice that for frequency larger than 10MHz, the variance $\Delta^2 X_-$ presents asymmetric behaviors reaching a situation in which the quantum correlations remains for pair 1 (P1) whereas for the pair 2 (P2) the noise becomes classical. Below the 10MHz bandwidth, both pairs present quantum correlation contributing for the noise compression observed in the symmetric basis (first row).

Now in the case of applying the dressing field at $\Delta_D=-1010$~MHz, one can observe that around 10MHz,  P2 loses the correlation whereas the P1 hold the correlations. Interestingly, figure (c) shows that now, if we use the dressing field at $\Delta_D=-1030$~MHz, the situation above is exchanged: P2 holds the quantum correlations and P1 becomes classical. Notice that the sum of the quadratures of each pair approaches  the classical limit, showing a dispersive profile, independently to the other pair. 
On the other hand,  at an analysis frequency of 15~MHz, the dressing field at $\Delta_D=-1030$~MHz in figure (c) generates a 6WM that transforms the P2 into a pair of quantum correlated side-bands.
Therefore,  the 6WM generated by the dressing field, which is tunable, can act in or out of phase inhibiting or generating the quantum correlations, independently in the pair of side-bands.
Is worth noting that, for the usual 4WM the asymmetry of squeezing  between the pairs can not be exchanged, as it is done with the dressing field.

Thus the dressing field effectively generates a 6WM process that superposed to the 4WM end up rotating the quadratures of modes $\hat{a}$ and $\hat{b}$, transforming correlations of amplitude into correlations of phase. This is an important result because the independent basis shows that the quantum correlation of the pairs can be controlled independently by the dressing field, whereas in the symmetric basis, the presence of the dressing field seems to affect both situations equally. The transition from one situation to another can be done by simply tuning the frequency of the dressing field with respect to the excited level. 
Such spectral control can, in principle, change the steering properties of the four modes. Unlike ref.\cite{Zhang20} in which the steering of cascade dressed PA-4WM is  studied, this approach of independent sidebands, which transforms a bipartite system into a four mode system,  and the effect of the dressing field on their quantum correlations, is compact and avoids extensive and complex schemes that require technical synchronization.

\begin{figure}[t!]
\centering
\begin{overpic}[width=14cm]{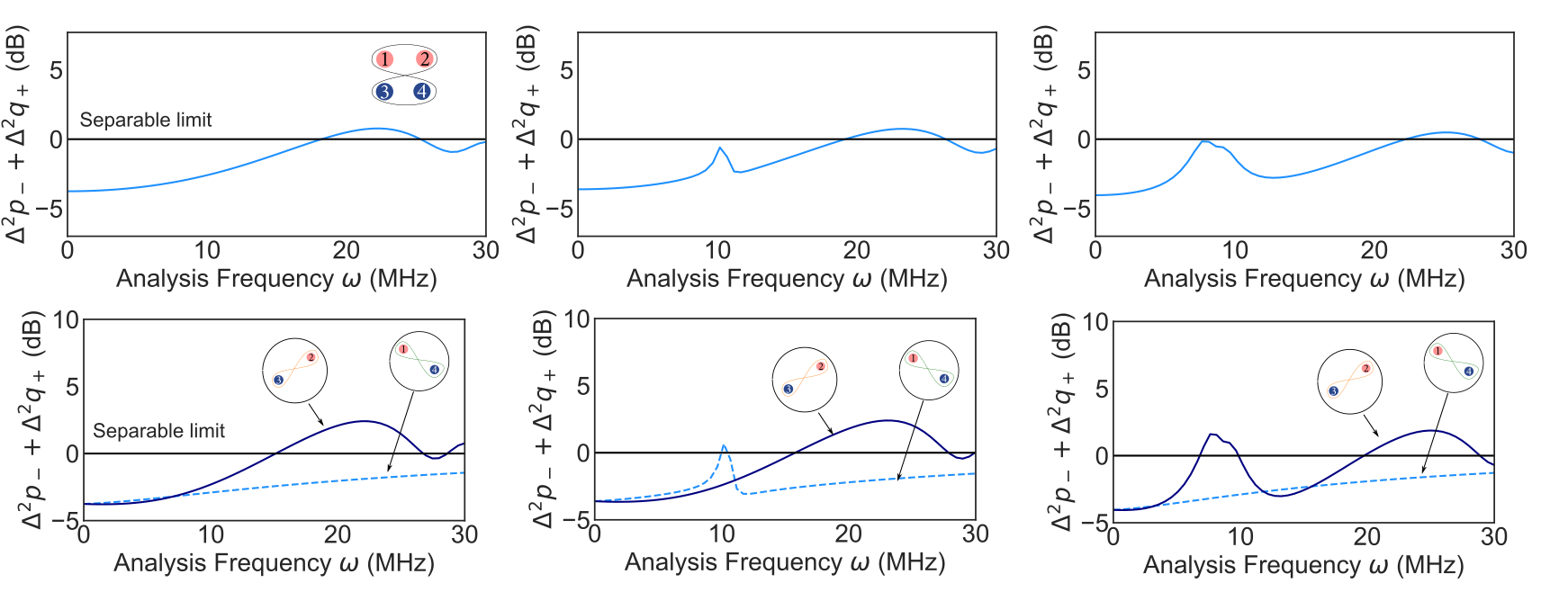}
\put(15,-4){(a)}
\put(50,-4){(b)}
\put(80,-4){(c)}
\end{overpic}
\vspace{0.5cm}
\caption{Spectrum of Duan criterion of entanglement for the symmetric basis (top) and independent basis (bottom). (a) 4WM without dressing field. (b) 4WM with interaction of the dressing field at $\Delta_D=-1010$~MHz and (c) at $\Delta_D=-1010$~MHz.}
\label{fig:4mode_duan_dressing}
\end{figure}

Now, looking to the Duan criterion the same control is preserved. 
Figure \ref{fig:4mode_duan_dressing} shows the same comparison done in Fig.~\ref{fig:4mode_dressing} for the Duan criterion.
Figure (a) shows in the top plot that the two fields are entangled within a range of 20~MHz. Looking into the independent basis, the plot at the bottom shows the asymmetry of entanglement in the spectrum. Meanwhile, Figs. (b) and (c) show the entanglement when the dressing field is coupled at $\Delta_D=-1010$~MHz and $\Delta_D=-1030$~MHz, respectively. Once again, around $\omega=10$~MHz, whilst the entanglement seems to be lost in the symmetric combined basis, in the case of independent basis, the entanglement is affected only in one of the pairs.

This tunabilty of quantum correlation could have great impact on the detection of quantum correlations. As it is discussed in ref~\cite{Felippe13}, such asymmetry would be unseen by standard homodyne detection, since only has access to the correlation in the first raw of Fig.~\ref{fig:4mode_dressing}. However, employing auto-homodyne detection with an optical resonator would be possible to detect such asymmetry accessing the correlation in the second raw of Fig.~\ref{fig:4mode_dressing}. 

\section{Conclusion}
We have presented a microscopical description of the 4WM process with the presence of an additional dressing field that leads into a superposition of 4WM and 6WM. We have shown that the model describes the increase of parametric amplification when the dressing field interacts with the atoms, showing consistent results with experimental observations. Furthermore, we have shown that the quantum correlations of the quadratures of the fields also increase due to the presence of the dressing field. In addition to that, the model allows to study the spectral control that the dressing field exerts on the quantum correlation of the twin beams of the 4WM process. In particular, we showed that the dressing fields can tune the entanglement of specific pairs of sidebands, which would be unseen by homodyne detection, but feasible detection with an optical resonator. This opens up a new possibility of spectral control of the steering among the frequency modes generated in the 4WM process.
%%%%%%%%%%%%%%%%%%%%%%% References %%%%%%%%%%%%%%%%%%%%%%%%%

\appendix
\section{\label{appendix}
Basis transformation for quadrature side-bands}
In order to transform the quadratures expressed in the symmetric combined basis $\delta\hat{\mathcal{Y}}(\omega)$ into independent sideband basis $\delta\hat{\mathcal{Y}}_\omega$, we employ the unitary transformation $\mathbf{L}_{ab}$ defined as
\begin{small}
\begin{align}
\mathbf{L}_{ab}=\frac{1}{2}\begin{bmatrix}
  1 & i &  0 & 0 & 1 & -i & 0 & 0\\
    -i & 1 & 0 & 0 & i & 1 & 0 & 0\\
    0 & 0 & 1 & i & 0 & 0 & 1 & -i \\
     0 & 0 & -i & 1 & 0 & 0 & i & 1\\
     1 & -i &  0 & 0 & 1 & i & 0 & 0\\
    i & 1 & 0 & 0 & -i & 1 & 0 & 0\\
    0 & 0 & 1 & -i & 0 & 0 & 1 & i \\
     0 & 0 & i & 1 & 0 & 0 & -i & 1\\
\end{bmatrix}.
\end{align}
\end{small}

Thus, we obtain the array of quadratures for both the beams $\hat{a}$ and $\hat{b}$ for the upper and lower  sidebands $\delta \hat{\mathbf{Y}}_{\pm\omega}$. 

% Add references with BibTeX or manually.
% \cite{Zhang:14,OSA,FORSTER2007,Dean2006,testthesis,Yelin:03,Masajada:13,codeexample}

%%%%%%%%%% If using BibTeX:
% \bibliography{sample}

%%%%%%%%%% If preparing manually:

\end{document}